\documentclass[12pt]{article}
\begin{document}

\title{\bf \Large Energy and Momentum in General Relativity}

\author{M. Sharif \thanks{E-mail: hasharif@yahoo.com}
\\ Department of Mathematics, University of the Punjab,\\ Quaid-e-Azam
Campus Lahore-54590, PAKISTAN.}

\date{}

\maketitle

\begin{abstract}
The energy and momentum for different cosmological models
using various prescriptions are evaluated. In particular, we have
focused our attention on the energy and momentum for gravitational
waves and discuss the results. It is concluded that there are
methods which can provide physically acceptable results.
\end{abstract}


\newpage

\section{Introduction}

The notion of energy has been one of the most thorny and important
problems in Einstein's theory of General Relativity (GR). There
have been many attempts [1,2,3] to get a well defined expression
for local or quasi-local energy and momentum. However, there is
still no generally accepted definition known. As a result,
different people have different points of view. Cooperstock [4]
argued that in GR, energy and momentum are localized in regions of
the non-vanishing energy and momentum tensor and consequently
gravitational waves are not carriers of energy and momentum in
vacuum. The gravitational waves, by definition, have zero
stress-energy tensor. Thus the existence of these waves was
questioned. However, the theory of GR indicates the existence of
gravitational waves as solutions of Einstein's field equations
[5]. Infact this problem arises because energy is not well defined
in GR.

The problem for gravitational waves was resolved by Ehlers and
Kundt [6], Pirani [7] and Weber and Wheeler [8] by considering a
sphere of test particles in the path of the waves. They showed
that these particles acquired a constant momentum from the waves.
Qadir and Sharif [9] presented an operational procedure, embodying
the same principle, to show that gravitational waves impart a
momentum. Rosen [10] investigated whether or not cylindrical
gravitational waves have energy and momentum. He used the
energy-momentum pseudo tensors of Einstein and Landau Lifshitz and
carried out calculations in cylindrical polar coordinates.
However, he arrived at the conclusion that the energy and momentum
density components vanish. These results supported the conjecture
of Scheidegger [11] that physical system cannot radiate
gravitational energy. Later, he pointed out [12] that the energy
and momentum densities turn out to be non-vanishing and reasonable
if the calculations are performed in Cartesian coordinates. Rosen
and Virbhadra [13] explicitly evaluated these quantities in the
Einstein's prescription by using Cartesian coordinates and found
them finite and well defined. Virbhadra [14] then used Tolman,
Landau-Lifshitz and Papapetrou's prescriptions to evaluate the
energy and momentum densities and found that the same results turn
out in all these prescriptions.

Energy and momentum density are usually defined by a second rank
tensor $T_a^b$. The conservation of energy and momentum are
described by the requirement that the tensor's divergence is zero.
However, in GR, the partial derivative in the usual conservation
equation $T^b_{a,b}=0$ is replaced by a covariant derivative.
$T_a^b$ then represents the energy and momentum of matter and all
non-gravitational fields and no longer satisfies $T^b_{a,b}=0$. A
contribution from the gravitational field must be added to obtain
an energy-momentum expression with zero divergence. Einstein
obtained such an expression and many others such as Landau and
Lifshitz, Papapetrou and Weinberg gave similar prescriptions
[15]. The expressions they gave are called energy-momentum
complexes because they can be expressed as a combination of
$T_a^b$ and a pseudotensor, which is interpreted to represent the
energy and momentum of the gravitational field. These complexes
have been heavily criticized because they are non-tensorial, i.e.
they are coordinate dependent. For the Einstein, Landau-Lifshitz,
Papapetrou, Weinberg (ELLPW) energy-momentum complexes, one gets
physically meaningful results only in {\it Cartesian coordinates}
[16-18]. Because of this drawback, many others, including Moller
[18], Komar [19] and Penrose [1], have proposed coordinate
independent definitions. Each of these, however, has its own
drawbacks [20,21].

In this paper we gather various results to show that
different prescriptions can provide the same result for different
cosmological models. Also, we shall see from the analysis that when
rotation is included, the problem becomes considerably complicated
and the results obtained may not be the same. This has been
explained by applying to a class of cylindrical gravitational
waves. The paper has been planned as follows. In the next section,
we shall describe different prescriptions to evaluate energy and
momentum densities. In section three, these methods will be
applied to different cosmological models. The section four
contains the evaluation of energy and momentum for gravitational
waves. Finally, we shall conclude the results.

\section{Various Prescriptions to Evaluate Energy and Momentum}

In this section we describe different methods to calculate energy
and momentum in GR.

\textbf{(i) Energy and Momentum in Einstein's Prescription}

The energy-momentum complex of Einstein [18,20] is given by
\begin{equation}
\Theta_a^b=\frac{1}{16\pi}H^{bc}_{a\;,c},
\end{equation}
where
\begin{equation}
H_a^{bc}=\frac{g_{ad}}{\sqrt{-g}}[-g(g^{bd}g^{ce}-g^{cd}g^{be})]_{,e},
\end{equation}
where Latin indices run from 0 to 3 and Greek indices from 1 to 3. $\Theta_0^0$
is the energy density, $\Theta_0^{\alpha}$ are the momentum
density components, and $\Theta_{\alpha}^0$ are the components of energy
current density. The Einstein energy-momentum satisfies the local
conservation laws
\begin{equation}
\frac{\partial\Theta_a^b}{\partial x^b}=0.
\end{equation}

\textbf{(ii) Energy and Momentum in Landau-Lifshitz Prescription}

The energy-momentum complex of Landau-Lifshitz [17] is given by
\begin{equation}
\L^{ab}=\frac{1}{16\pi}S^{abcd}_{,cd},
\end{equation}
where
\begin{equation}
S^{abcd}=-g[g^{ab}g^{cd}-g^{ac}g^{bd}].
\end{equation}
$L^{ab}$ is symmetric in its indices. $L^{00}$ is the energy density,
$L^{0\alpha}$ are the momentum (energy current) density
components. $S^{abcd}$ has symmetries of the Riemann curvature tensor.
The energy-momentum complex of Landau and Lifshitz satisfies the local
conservation laws
\begin{equation}
\frac{\partial L^{ab}}{\partial x^b}=0,
\end{equation}
where
\begin{equation}
L^{ab}=-g(T^{ab}+t^{ab}).
\end{equation}
$g$ is the determinant of the metric tensor $g_{ab}$, $T^{ab}$ is
the energy-momentum tensor of the matter and all non-gravitational
fields, and $t^{ab}$ is known as Landau-Lifshitz energy-momentum pseudo
tensor. Thus the locally conserved quantity $L^{ab}$ contains contributions
from the matter, non-gravitational fields and gravitational fields.

\textbf{(iii) Energy and Momentum in Papapetrou's Prescription}

The energy-momentum complex of Papapetrou [22] is given by
\begin{equation}
\Omega^{ab}=\frac{1}{16\pi}N^{abcd}_{,cd},
\end{equation}
where
\begin{equation}
N^{abcd}=\sqrt{-g}[g^{ab}\eta^{cd}-g^{ac}g^{bd}+g^{cd}\eta^{ab}-g^{bd}g^{ac}],
\end{equation}
and $\eta^{ab}$ is the Minkowski spacetime. $\Omega^{00}$ and
$\Omega^{\alpha 0}$ are the energy and momentum density components respectively.
The Papapetrou energy-momentum complex satisfies the local conservation laws
\begin{equation}
\frac{\partial\Omega^{ab}}{\partial x^b}=0.
\end{equation}

\textbf{(iv) Energy and Momentum in Weinberg Prescription}

The energy-momentum complex of Weinberg [23] is given by
\begin{equation}
W^{ab}=\frac{1}{16\pi}\Delta^{abc}_{,c},
\end{equation}
where
\begin{equation}
\Delta^{abc}=\frac{\partial h_e^e}{\partial x_a}\eta^{bc}-
\frac{\partial h_e^e}{\partial x_b}\eta^{ac}-
\frac{\partial h^{ea}}{\partial x^e}\eta^{bc}+
\frac{\partial h^{eb}}{\partial x^e}\eta^{ac}+
\frac{\partial h^{ac}}{\partial x_b}-\frac{\partial h^{bc}}{\partial x_a}
\end{equation}
and
\begin{equation}
h_{ab}=g_{ab}-\eta_{ab}.
\end{equation}
The indices on $h_{ab}$ or
$\frac{\partial}{\partial x_a}$ are raised or lowered with the
help of $\eta$'s. The Weinberg energy-momentum complex $W^{ab}$
contains contributions from the matter, non-gravitational and
gravitational fields, and satisfies the local conservation laws
\begin{equation}
\frac{\partial W^{ab}}{\partial x^b}=0.
\end{equation}
$W^{00}$ and $W^{\alpha 0}$ are the energy and momentum density components
respectively.

\textbf{(v) Energy and Momentum in M$\ddot{o}$ller Prescription}

The energy-momentum complex of Moller [18] is given by
\begin{equation}
M_a^b=\frac{1}{8\pi}\chi^{bc}_{a,c},
\end{equation}
satisfying the local conservation laws:
\begin{equation}
\frac{\partial M_a^b}{\partial x^b}=0,
\end{equation}
where the antisymmetric superpotential $\chi_a^{bc}$ is
\begin{equation}
\chi_a^{bc}=\sqrt{-g}[g_{ad,e}-g_{ae,d}]g^{be}g^{cd}.
\end{equation}
$M_0^0$ is the energy density and $M_{\alpha}^0$ are the momentum density
components.

\textbf{(vi) Energy and Momentum in Qadir-Sharif's Prescription}

The energy-momentum complex of Qadir-Sharif [9] is given by
\begin{equation}
p_a=\int F_a,
\end{equation}
where
\begin{equation}
F_0=m[\{\ln(A/\sqrt{g_{00}})\}_{,0}-g_{\alpha\beta,0}g^{\alpha\beta}_{,0}/4A],\quad
F_i=m(\ln\sqrt{g_{00}})_{,\alpha},
\end{equation}
and $A=(\ln\sqrt{-g})_{,0}$,\quad $g=det(g_{\alpha\beta})$. This force formula
depends on the choice of frame, which is not uniquely fixed. The quantity,
whose proper time derivative is $F_a$, is called the momentum four-vector
for the test particle. The spatial components of $p_a$ give the momentum
imparted to test particles as defined in the preferred frame (in which $g_{0\alpha}=0$).

\section{Application to Various Cosmological Models}

\textbf{(i) Bianchi Type I Universes}

The Bianchi type I spacetimes are expressed by the line element
\begin{equation}
ds^2=dt^2-e^{2l}dx^2-e^{2m}dy^2-e^{2n}dz^2,
\end{equation}
where $l,m,n$ are functions of $t$ alone. Using ELLPW prescriptions,
it turns out that energy-momentum distribution is zero. This supports
the viewpoint of Tyron [24].

\textbf{(ii) Axially Symmetric Scalar Field}

It is well known that the Kaluza-Klein and the superstring theories
predict the scalar fields as a fundamental interaction in Physics.
Scalar fields are fundamental components of the Brans-Dicke theory
and of the inflationary models. Also, they are a good candidate for
the dark matter in spiral galaxies. Because they interact very weakly
with matter we have never seen one but many of the theories containing
scalar fields are in good concordance with measurements in weak gravitational
fields. Also, we expect that they can play an important role in strong
gravitational fields like at the origin of the universe or in pulsars
or black holes. The metric that we consider [25,26] is an axially
symmetric solution to the field equations derived from the action for
gravity minimally coupled to a scalar field. The solution is

\begin{equation}
ds^2=(1-\frac{2M}{r})dt^2-\frac{e^{2k_a}}{1-\frac{2M}{r}}dr^2-r^2(e^{2k_a}d\theta^2
+\sin^2\theta d\phi^2),
\end{equation}
with
\begin{equation}
e^{2k_a}=(1+\frac{M^2\sin^2\theta}{r^2(1-\frac{2M}{r})})^\frac{-1}{a^2},\quad
\phi=\frac{1}{2a}\ln(1-\frac{2M}{r}),
\end{equation}
where $a$ is a constant of integration and $\phi$ is the scalar field.
This solution is one of the new classes of solutions to the Einstein-Maxwell
theory non-minimally coupled to a dilatonic [26]. The metric given by Eq.(21)
is almost spherically symmetric and represents a gravitational body
(gravitational monopole) with scalar field. The scalar field deforms the
spherically symmetry. We observe that when $a\rightarrow \infty$ we recover
the Schwarzschild solution.

When we apply M$\ddot{o}$ller's prescription, it yields
\begin{equation}
E=M.
\end{equation}
Thus the energy distribution is given by the mass $M$. In the case of the
Schwarzschild metric we obtain the same result.

\textbf{(iii) Charged Regular Black Hole}

The Reissner-Nordstrom (RN) metric is the only static and asymptotically
flat solution of the Einstein-Maxwell equations and it represents an
electrically charged black hole. The metric is given by
\begin{equation}
ds^2=A(r)dt^2-B(r)dr^2-r^2(d\theta^2+\sin^2\theta d\phi^2),
\end{equation}
where
\begin{equation}
A(r)=B^{-1}(r)=1+\frac{2M}{r}+\frac{q^2}{r^2}
\end{equation}
and $q$ and $M$ are the electric charge and the mass of the black hole
respectively.

A solution to the coupled system of the Einstein field equations
of the nonlinear electrodynamics was recently given by E. Ayon-Beato
and A. Garcia (ABG) [27]. This solution represents a regular black
hole with mass $M$ and electric charge $q$ and avoids thus the
singularity problem. Also, the metric asymptotically behaves as the
RN solution. The usual singularity of the RN solution,
at $r=0$, has been smoothed out and now it simply corresponds to the
origin of the spherical coordinates. The line element is given by (24) with
\begin{equation}
A(r)=B^{-1}(r)=1-\frac{2M}{r}(1-\tanh(\frac{q^2}{2Mr})).
\end{equation}
If the electric charge vanishes we reach the Schwarzschild solution.
At large distances (26) resembles to the RN solution and can be written

\begin{equation}
A(r)=B^{-1}(r)=1+\frac{2M}{r}+\frac{q^2}{r^2}+\frac{q^6}{12M^2r^4}+O(\frac{1}{r^6}).
\end{equation}

Using Einstein's prescription, we get the energy distribution of the
ABG black hole given by
\begin{equation}
E(r)=M-\frac{q^2}{2r}+\frac{q^6}{24M^2r^3}-\frac{q^{10}}{240M^4r^5}+O(\frac{1}{r^6}).
\end{equation}
This can also be written as
\begin{equation}
E(r)=E_{RN}(r)+\frac{q^6}{24M^2r^3}-\frac{q^{10}}{240M^4r^5}+O(\frac{1}{r^6}).
\end{equation}
It follows that if $q=0$ we have the energy of a Schwarzschild black hole.
The M$\ddot{o}$ller's prescription gives
\begin{equation}
E(r)=M-\frac{q^2}{r}+\frac{q^6}{6M^2r^3}-\frac{q^10}{40M^4r^5}+O(\frac{1}{r^6}).
\end{equation}
From Eq.(28) it results that in the Einstein prescription the first
two terms in the expression of the energy correspond to the Penrose
quasi-local mass definition evaluated by Tod [1,28]. The M$\ddot{o}$ller's prescription
provides in the expression of the energy (30) a term $M-\frac{q^2}{r}$
which agrees with the Komar [19] prescription.

\textbf{(iv) Kerr-Newmann Metric}

The stationary axially symmetric and asymptotically flat Kerr-Newmann
(KN) solution is the most general black hole solution to the Einstein-Maxwell
equations. This describes the exterior gravitational and electromagnetic
field of a charged rotating object. The KN metric in
Boyer-Lindquist coordinates $(t,\rho,\theta,\phi)$ is expressed by
the line element
\begin{equation}
ds^2=\frac{\Delta}{r^2_0}[dt-a\sin^2\theta d\phi]^2-\frac{\sin^2\theta}{r^2_0}
[(\rho^2+a^2)d\phi-adt]^2-\frac{r^2_0}{\Delta}d\rho^2-r^2_0d\theta^2,
\end{equation}
where $\Delta=\rho^2-2M\rho+q^2+a^2$ and $r^2_0=\rho^2+a^2\cos^2\theta$.
$M, q$ and $a$ are respectively mass, electric charge and rotation parameters.
Aguirregabiria et al. [29] studied the energy-momentum complexes of ELLPW for
the KN metric. They showed that these definitions give the same results for
the energy and energy current densities. This is given as
\begin{equation}
E_{ELLPW}=M-\frac{q^2}{4\rho}[1+\frac{(a^2+\rho^2)}{a\rho}\arctan(\frac{a}{\rho})].
\end{equation}
Using M$\ddot{o}$ller's prescription, we have
\begin{equation}
E_{M\ddot{o}l}=M-\frac{q^2}{2\rho}[1+\frac{(a^2+\rho^2)}{a\rho}\arctan(\frac{a}{\rho})].
\end{equation}
The result of M$\ddot{o}$ller's agrees with the energy distribution
obtained by Cohen and de Felice [30] in Komar's prescription.
The second term of the energy distribution differs by a factor
of two from that computed by Aguirregabiria et al. using ELLPW complex.
The total energy ($\rho\rightarrow\infty$ in all these energy expressions)
give the same result $M$.

\textbf{(v) Melvin's Magnetic Universe}

Melvin [31] obtained an axially symmetric electrovac solution ($J^a=0$)
describing the Schwarzschild black hole in Melvin's magnetic universe.
The spacetime is
\begin{equation}
ds^2=\Lambda^2[(1-\frac{2M}{r})dt^2-(1-\frac{2M}{r})^{-1}dr^2-r^2d\theta^2]
-\Lambda^{-2}r^2\sin^2\theta d\phi^2
\end{equation}
and the Cartan components of the magnetic field are
\begin{equation}
H_r=\Lambda^{-2}B_0\cos\theta,\quad
H_{\theta}=-\Lambda^{-2}B_0(1-2M/r)^{1/2}\sin\theta,
\end{equation}
where
\begin{equation}
\Lambda=1+\frac{1}{4}B^2_0r^2\sin^2\theta.
\end{equation}
$M$ and $B_0$ are constants. The ELLP energy distribution gives
\begin{equation}
E=Mc^2+\frac{1}{6}B^2_0r^3+\frac{1}{20}\frac{G}{c^4}B^4_0r^5
+\frac{1}{140}\frac{G^2}{c^8}B^6_0r^7+\frac{1}{2520}\frac{G^3}{c^12}B^8_0r^9.
\end{equation}
The above result can be expressed in geometrized units
(gravitational constant $G=1$ and the speed of light in
vacuum $c=1$) as follows
\begin{equation}
E=M+\frac{1}{6}B^2_0r^3+\frac{1}{20}B^4_0r^5+\frac{1}{140}B^6_0r^7+
\frac{1}{2520}B^8_0r^9.
\end{equation}
The first term $Mc^2$ is the rest-mass energy of the Schwarzschild black hole,
the second term $\frac{1}{6}B^2_0r^3$ is the well-known classical value of the
energy of the magnetic field under consideration, and rest of the terms are
general relativistic corrections. For very large $B_0r$, the general
relativistic contribution dominates over the classical value for the
magnetic field energy.

\section{Application to Gravitational Waves}

\textbf{(i) Plane Gravitational Waves}

The metric for the plane-fronted gravitational waves is [5,32]
\begin{equation}
ds^2=dt^2-dx^2-L^2(t,x)[exp\{2\beta(t,x)\}dy^2+exp\{-2\beta(t,x)\}dz^2],
\end{equation}
where $L$ and $\beta$ are arbitrary functions subject to the vacuum
Einstein equations
\begin{equation}
L_{,\alpha\alpha}+L\beta^2_{,\alpha}=0,\quad \alpha=0,1.
\end{equation}
Since $L$ and $\beta$ are functions of $u=t-x$, Eqs.(40) reduce to the single equation
\begin{equation}
L_{uu}+L\beta^2_{u}=0.
\end{equation}
Using Qadir-Sharif procedure, we obtain
\begin{equation}
F_0=-m(\ddot{L}+\dot{\beta}^2)L/\dot{L}=0,\quad F_i=0,
\end{equation}
where a dot denotes differentiation with respect to $t$. Consequently the
momentum four-vector becomes $p_a=constant$. Thus there is a constant
energy and momentum. The constant, here, determines the strength of the wave.
This exactly coincides with the Ehlers-Kundt method in which they
demonstrate that the test particles acquire a constant momentum and
hence a constant energy, from the plane gravitational waves.

\textbf{(ii) Cylindrical Gravitational Waves}

To describe cylindrical gravitational waves one uses cylindrical
polar coordinates $(\rho,\theta,\phi)$ and the time $t$, and one
takes the line element in the form [8]
\begin{equation}
ds^2=e^{2\gamma-2\psi}(dt^2-d\rho^2)-\rho^2e^{-2\psi}d\phi^2-e^{2\psi}dz^2,
\end{equation}
where $\gamma=\gamma(\rho,t)$, $\psi=\psi(\rho,t)$. To satisfy the
Einstein field equations for empty space one takes
\begin{equation}
\psi_{tt}-\psi_{\rho\rho}-\frac{1}{\rho}\psi_{\rho}=0,
\end{equation}
\begin{equation}
\gamma_t=2\rho\psi_{\rho}\psi_{t},\quad \gamma_{\rho}=\rho(\psi^2_{\rho}+\psi_{t}),
\end{equation}
where subscript denote partial derivatives.
Using ELLP prescription, we obtain the energy-momentum densities given by
\begin{equation}
\Theta_0^0=\frac{1}{8\pi}e^{2\gamma}(\psi_\rho^2+\psi_t^2),
\end{equation}
\begin{equation}
\Theta_1^0=\frac{1}{4\pi\rho}x\psi_\rho\psi_t,
\end{equation}
\begin{equation}
\Theta_2^0=\frac{1}{4\pi\rho}y\psi_\rho\psi_t,
\end{equation}\begin{equation}
\Theta_0^1=-e^{2\gamma}\Theta_1^0,
\end{equation}
\begin{equation}
\Theta_0^2=-e^{2\gamma}\Theta_2^0,
\end{equation}
\begin{equation}
\Theta_0^3=\Theta_3^0=0.
\end{equation}
The energy density of the cylindrical gravitational waves is finite
and positive definite, and the momentum density components reflect the
symmetry of the spacetime.

\textbf{(iii) A Class of Rotating Cylindrical Gravitational Waves}

A class of solutions of the gravitational field equations
describing vacuum spacetimes outside rotating cylindrical sources
is given by the line element of the form [33]
\begin{equation}
ds^2=e^{2\gamma-2\psi}(dt^2-d\rho^2)-\mu^2e^{-2\psi}(\omega
dt+d\phi)^2-e^{2\psi}dz^2,
\end{equation}
in the cylindrical coordinates $(\rho,\phi,z)$. Here the metric
functions $\gamma,\mu,\psi$ and $\omega$ depend on the coordinates
$t$ and $\rho$ only. When $\omega=0$, the metric represents spacetimes
without rotation, in which the polarization of gravitational waves
has only one degree of freedom and the direction of polarization
is fixed [5]. It is to be noticed that if we take $\omega=0$ and
$\mu=\rho$, the above metric reduces to a special case of
cylindrical gravitational waves [8]. Einstein's vacuum field
equations for the metric form (52) are given by
\begin{equation}
(\mu\psi_v)_u+(\mu\psi_u)_v=0,
\end{equation}
\begin{equation}
\mu_{uv}-\frac{l^2}{8}\mu^{-3}e^{2\gamma}=0,
\end{equation}
\begin{equation}
\omega_v-\omega_u=l\mu^{-3}e^{2\gamma},
\end{equation}
\begin{equation}
\gamma_u=\frac{1}{2\mu_u}(\mu_{uu}+2\mu\psi^2_u),
\end{equation}
\begin{equation}
\gamma_v=\frac{1}{2\mu_v}(\mu_{vv}+2\mu\psi^2_v),
\end{equation}
where $\psi_u=\frac{\partial\psi}{\partial u}$, etc. The
subscripts $u=t-\rho$ and $v=t+\rho$ are retarded and advanced
times respectively. Here $l$ is a constant length characteristic
of the rotation of the system which is positive and is
specifically attributed with rotating gravitational waves. For
$l=0$ we have $\omega=\omega(t)$ from Eq.(55) and
$\mu_{tt}=\mu_{\rho\rho}$ from Eq.(54). A simple transformation to
a rotating frame reduces the waves to non-rotating generalized
Beck spacetimes which have been studied by many authors [32,34,35].

The energy and momentum densities in Einstein's prescription are
\begin{eqnarray}
\Theta_0^0=\frac{1}{16\pi\mu^2\rho^3}[\mu^2(-\mu+\mu_\rho\rho-2\mu_{\rho\rho}\rho^2
-\mu\omega^2\rho^2-2\mu\omega\omega_\rho\rho^3-\mu_\rho\omega^2\rho^3)\\\nonumber
+\rho^2(\mu+2\mu\gamma_\rho\rho-\mu_\rho\rho)e^{2\gamma}-\mu^4\rho^2(\mu\omega_\rho^2
+\mu\omega\omega_{\rho\rho}\\\nonumber
-2\mu\omega\omega_\rho\gamma_\rho+3\mu_\rho\omega\omega_\rho)e^{-2\gamma}],
\end{eqnarray}
\begin{eqnarray}
\Theta_1^0=\frac{1}{16\pi\rho^5}[2\mu\rho^2\dot{\gamma}x-6\dot{\mu}x^3-2\dot{\mu}\rho^2x
-2\dot{\mu_\rho}\rho x^3 -\mu\omega^2\rho y\\\nonumber -\mu^2\rho
y(\mu\omega_\rho
-\mu\omega_{\rho\rho}\rho+2\mu\omega_\rho\gamma_\rho\rho-3\mu\mu_\rho\omega_\rho\rho)
e^{-2\gamma}],
\end{eqnarray}
\begin{eqnarray}
\Theta_2^0=\frac{1}{16\pi\rho^4}[2\mu\rho\dot{\gamma}y-2\dot{\mu_\rho}\rho^2y
+\mu\omega\rho
x-\mu\omega_{\rho}\rho^2x-\mu_\rho\omega\rho^2x\\\nonumber
+\mu^2x(\mu\omega_\rho-\mu\omega_{\rho\rho}\rho
+2\mu\omega_\rho\gamma_\rho\rho-3\mu_\rho\omega_\rho\rho)e^{-2\gamma}],
\end{eqnarray}
\begin{eqnarray}
\Theta_0^1=\frac{1}{16\pi\mu^2\rho^3}[\mu^2(-\dot{\mu}x+2\dot{\mu_\rho}\rho
x+2\mu\omega\dot{\omega}\rho^2x-2\mu\omega\ddot{\gamma}\rho^2y
+2\mu\omega_\rho\gamma_\rho\rho^2y\\\nonumber
-\mu\omega_{\rho\rho}\rho^2y+2\mu\omega\gamma_{\rho\rho}\rho^2y
+\dot{\mu}\omega^2\rho^2x+2\mu_\rho\omega\gamma_\rho\rho^2y
-3\mu_\rho\omega_\rho\rho^2y\\\nonumber
-2\mu_{\rho\rho}\omega\rho^2y)-\rho^2x(2\mu\dot{\gamma}-\dot{\mu})e^{2\gamma}
+\mu^4\rho x(\mu\dot{\omega}\omega_\rho\\\nonumber
+\mu\omega\dot{\omega_\rho}
-3\mu\omega\omega_\rho\dot{\gamma}+3\dot{\mu}\omega\omega_\rho)e^{-2\gamma}],
\end{eqnarray}
\begin{eqnarray}
\Theta_0^2=\frac{1}{16\pi\mu^2\rho^3}[\mu^2(2\dot{\mu_\rho}\rho y
+\mu\omega_{\rho\rho}\rho^2y-2\mu\omega_\rho\gamma_\rho\rho^2y
-2\mu\omega\gamma_{\rho\rho}\rho^2y+2\mu\omega\dot{\omega}\rho^2y\\\nonumber
+2\mu\dot{\omega}\dot{\gamma}\rho^2x
+2\mu\omega\ddot{\gamma}\rho^2x
+\dot{\mu}\omega^2\rho^2y+2\dot{\mu}\omega\dot{\gamma}\rho^2
x+3\mu_\rho\omega_\rho\rho^2y\\\nonumber
-2\mu_\rho\omega\gamma_\rho\rho^2y
+2\mu_{\rho\rho}\omega\rho^2y)-\rho^2y(2\mu\dot{\gamma}-\dot{\mu})e^{2\gamma}
+\mu^4\rho
y(\mu\dot{\omega}\omega_\rho\\\nonumber+\mu\omega\dot{\omega_\rho}
-2\mu\omega\omega_\rho\dot{\gamma}
+3\dot{\mu}\omega\omega_\rho)e^{-2\gamma}],
\end{eqnarray}
\begin{equation}
\Theta_0^3=\Theta_3^0=0.
\end{equation}
Now for $\omega=0$ and $\mu=\rho$, Eqs.(58)-(62) become
the energy and momentum densities of cylindrical
gravitational waves given by Rosen and Virbhadra [13].

The energy and momentum density components in Papapetrou's prescription is given by
\begin{eqnarray}
\Omega^{00}=\frac{1}{16\pi\mu^2\rho^3}[\mu^2(-\mu+\mu_\rho\rho-2\mu_{\rho\rho}\rho^2
-\mu\omega^2\rho^2-2\mu\omega\omega_\rho\rho^3-\mu_\rho\omega^2\rho^3)\\\nonumber
+(\mu\rho^2+2\mu\gamma_\rho\rho^3-\mu_\rho\rho^3)e^{2\gamma}],
\end{eqnarray}
\begin{eqnarray}
\Omega^{01}=\frac{1}{16\pi\mu^2\rho^3}[\mu^2(-\dot{\mu}x+2\dot{\mu_\rho}\rho
x+\mu\omega y-\mu\omega_\rho\rho
y-\mu\omega_{\rho\rho}\rho^2y\\\nonumber
+2\mu\omega\dot{\omega}\rho^2x+\dot{\mu}\omega^2\rho^2x-\mu_\rho\omega\rho
y-2\mu_\rho\omega_\rho\rho^2y\\\nonumber
-\mu_{\rho\rho}\omega\rho^2y)-\rho^2x(2\mu\dot{\gamma}-\dot{\mu})e^{2\gamma}],
\end{eqnarray}
\begin{eqnarray}
\Omega^{02}=\frac{1}{16\pi\mu^2\rho^3}[\mu^2(\dot{\mu}y+2\dot{\mu_\rho}\rho
y+2\mu\omega\dot{\omega}\rho^2y+\dot{\mu}\omega^2\rho^2y-\mu\omega
x\\\nonumber +\mu\omega_\rho\rho x+\mu\omega_{\rho\rho}\rho^2x
+\mu_\rho\omega \rho x+2\mu_\rho\omega_\rho\rho^2x\\\nonumber
+\mu_{\rho\rho}\omega\rho^2x)-\rho^2
y(2\mu\dot{\gamma}-\dot{\mu})e^{2\gamma}],
\end{eqnarray}
\begin{equation}
\Omega^{03}=\Omega^{30}=0.
\end{equation}
We see that for $\omega=0$ and $\mu=\rho$, Eqs.(64)-(66) yield the same result
as given by Virbhadra [13].

\textbf{(iv) Spherical Gravitational Waves}

The gravitational waves with spherical wavefronts are given by the
line element of the form [36]
\begin{equation}
ds^2=e^{-M}(dt^2-d\rho^2)-e^{-U}(e^Vdz^2+e^{-V}d\phi^2),
\end{equation}
where the metric functions $U,V$ and $M$ depend on the coordinates
$t$ and $\rho$ only. Einstein's vacuum field equations imply that
$e^{-U}$ satisfies the wave equation
\begin{equation}
(e^{-U})_{tt}-(e^{-U})_{\rho\rho}=0,
\end{equation}
and that $V$ satisfies the linear equation
\begin{equation}
V_{tt}-U_tV_t-V_{\rho\rho}+U_{\rho}V_{\rho}=0.
\end{equation}
The remaining equations for $M$ are
\begin{equation}
U_{tt}-U_{\rho\rho}=\frac 12
(U_t^2+U_{\rho}^2+V_t^2+V_{\rho}^2)-U_tM_t-U_{\rho} M_{\rho}=0,
\end{equation}
\begin{equation}
2U_{t\rho}=U_tU_{\rho}-U_tM_{\rho}-U_{\rho}M_t+V_tV_{\rho}.
\end{equation}
It is well known that, if Eqs.(69) and (70) are satisfied, the
Eqs.(71) and (72) are automatically integrable.

Using Qadir-Sharif prescription, we obtain
\begin{equation}
F_0=m[\dot{U}+\frac{\ddot{M}+2\ddot{U}}{\dot{M}+2\dot{U}}-
\frac{3\dot{U}^2+\dot{V}^2}{\dot{M}+2\dot{U}}],\quad
F_1=-m\frac{M'}{2},\quad F_2=0=F_3.
\end{equation}
The corresponding four-vector momentum will become
\begin{equation}
p_0=m[U+\ln(\dot{M}+2\dot{U})-\int
\frac{3\dot{U}^2+\dot{V}^2}{\dot{M}+2\dot{U}}dt]+f_1(\rho),
\end{equation}
\begin{equation}
p_1=-\frac{m}{2}\int M'dt+f_2(\rho),\quad p_2=constant=p_3.
\end{equation}
where dot denotes differentiation with respect to time and prime
with respect to $\rho$, $f_1$ and $f_2$ are arbitrary functions of
$\rho$. Eqs.(74) and (75) provide the general expression of the
momentum four-vector for the gravitational waves with spherical
wavefronts. As we are interested in evaluating the momentum
imparted  by gravitational waves we need to calculate the term
$p_1$. For this purpose, we require the value of $M$.

The background region ($t<\rho$, Minkowski) is described by the
solution \\ $U=-\ln t-\ln\rho,\quad V=\ln t-\ln\rho$ and $M=0$.
Substituting these values in Eqs.(74) and (75), we have
\begin{equation}
p_0=m\ln(-2/\rho)+f_1(\rho),\quad p_{\alpha}=constant.
\end{equation}
The quantity $p_0$ can be made zero by choosing
$f_1(\rho)=-m\ln(-2/\rho)$ and the momentum term $p_{\alpha}$ will be
zero for a particular choice of an arbitrary constant as zero.
Thus the four-vector momentum vanishes in the background region
(Minkowski) as was expected.

The solution on the wavefront ($t=\rho$) can be written in the
form\\ $U=-2\ln t,\quad V=0,\quad M=0$. Using these values in
Eqs.(74) and (75), it follows that
\begin{equation}
p_0=m\ln(-4)+f_1(\rho),\quad p_{\alpha}=constant.
\end{equation}
We see that the momentum turns out to be constant which can be
made zero if we choose constant as zero.

The solution in the wave region ($t>\rho$) can be found by solving
Eqs.(69) and (70) and is given in the form [35,36]
\begin{equation}
U=-\ln t-\ln\rho,\quad V=\ln t-\ln\rho+\tilde{V}(t,\rho).
\end{equation}
The case of a single component gives
\begin{equation}
\tilde{V}(t,\rho)=a_k(t\rho)^kH_k(\frac{t^2+\rho^2}{2t\rho})
\end{equation}
for some $k\geq\frac 12$ and constant $a_k$. In this case, $M$ is
given by
\begin{equation}
M=\frac{1}{2k}a_k(t^2-\rho^2)(t\rho)^{k-1}H_{k-1}-\frac{1}{2k}
(t\rho)^{2k}a^2_k[k^2H^2_k-\frac{(t^2-\rho^2)^2}{4t^2\rho^2}H^2_{k-1}].
\end{equation}
Notice that the dimension of $a_k$ is $L^{-2k}$. For the purpose of
simplicity, we take a special case when $k=1$ for which $M$ takes
the form
\begin{equation}
M=\frac{1}{2}a_1(t^2-\rho^2)H_0-\frac{1}{2}(t\rho)^2a^2_1[H^2_1
-\frac{(t^2-\rho^2)^2}{4t^2\rho^2}H^2_0],
\end{equation}
where $H_0=\ln(t/\rho),\quad H_1=\frac
12[(t/\rho+\rho/t)\ln(t/\rho)-(t/\rho-\rho/t)]$. After taking
derivative of $M$ with respect to $\rho$, we substitute it in
Eq.(75) and after a tedious integration, we obtain
\begin{eqnarray}
p_1=m\frac{a_1}{4}[\frac 23(t^3/\rho)-6t\rho+4t\rho\ln(t/\rho)
+a_1\{\frac{1}{5}(t^5/\rho)+\frac{2}{27}t^3\rho+5t\rho^3\\\nonumber
-4(\frac{5}{9}t^3\rho+t\rho^3)\ln(t/\rho)+\frac{4}{3}t^3\rho(\ln(t/\rho))^2\}]
+f_2(\rho).
\end{eqnarray}
This gives the momentum imparted to test particles by
gravitational waves with spherical wavefronts. The quantity $p_1$
can be made zero for the $t\rightarrow 0$ limit by choosing
$f_2=0$. However, it immediately indicates the presence of
singularity when $\rho=0$ and this singularity at $\rho=0$ acts as
a source of the gravitational waves inside the wave region. This
coincides with the result evaluated by using M$\ddot{o}$ller's
prescription [18]. This is a physically reasonable expression for
the momentum imparted by gravitational waves. The interpretation
of $p_0$ in the $e\psi N$ formalism is given elsewhere [37]. It
can also be shown that, near the wavefront as $t\rightarrow\rho$
\begin{equation}
\tilde{V}\sim a_k(t+\rho)^{-1}(t-\rho)^{1+2k},\quad M\sim
(\frac{1+2k}{2k})a_k(t-\rho)^{2k}.
\end{equation}
Using Eqs.(75) and (83), it follows that
\begin{equation}
p_1=\frac{1}{4k}m(1+2k)a_k(t-\rho)^{2k}+f_2(\rho).
\end{equation}
This gives the momentum near the spherical wavefront. We see that
for a particular choice of $f_2$, it reduces to the momentum
expression given by Eq.(77) on the wavefront. We remark that our
results exactly coincide with those evaluated by using
M$\ddot{o}$ller's prescription for the background region and on
the wavefront. For the wave region, these two can be equated for a
particular choice of an arbitrary function $f_2$. We have seen
that in all the three cases we obtain a physically reasonable
expression for the momentum.

\section{Discussion}

It is usually believed that different energy-momentum complexes
could give different results for a given geometry . Keeping this
point in mind, we have applied various prescriptions to different
cosmological models. We have also extended this analysis to gravitational
waves. For many spacetimes, it is found that various methods could give
the same result which is physically well-defined.

The Bianchi type I metric gives zero energy-distribution in
ELLPW prescriptions. For the axially symmetric scalar field, using
M$\ddot{o}$ller's method, the energy distribution becomes the mass $M$
which coincides with the Schwarzschild metric. We also obtain the
physically interesting results for the charged regular black hole,
KN metric and Melvin's magnetic universe as given in the analysis.

Further, we have evaluated energy and momentum distribution for gravitational
waves by different prescriptions. The Ehler-Kundt method gives the physically
reasonable result that plane gravitational waves impart a constant energy
and momentum to test particles in their path. However, it does not provide
a simple formula that can be applied to other cases. We have obtained the same
result using Qadir-Sharif formalism. For cylindrical gravitational waves,
using ELLP energy-momentum distribution, we have obtained the similar
and physically acceptable result. We have also calculated the
energy-momentum distribution for a class of cylindrical gravitational
waves using EP prescriptions. It can be
seen that the energy and momentum densities for a class of rotating
gravitational waves are finite and well-defined in both the
prescriptions. It follows from Eqs.(58-63) and (64-67) that though
the energy-momentum complexes of Einstein and Papapetrou are not
exactly the same but are similar upto certain terms. However, it
is interesting to note that both the results reduce to the same
energy and momentum densities of a special case of cylindrical gravitational
waves as given in [13,14].

Finally, we have applied the Qadir-Sharif prescription to spherical
gravitational waves. It is interesting to note that this provide physical
acceptable result. This result supports the result evaluated by using
M$\ddot{o}$ller's prescription. We can conclude that different
prescriptions can provide the same meaningful result.

\newpage

\begin{description}
\item  {\bf Acknowledgment}
\end{description}

The author would like to thank MOST for funding this work.

\vspace{2cm}

{\bf \large References}

\begin{description}

\item{[1]} Penrose, R.: Proc. Roy. Soc. London {\bf
A381}(1982)53.
\item{[2]} Misner, C.W. and Sharp, D.H. Phys. Rev. {\bf
136}(1964)B571.

\item{[3]} Bondi, H.: Proc. Roy. Soc. London {\bf A427}(1990)249.

\item{[4]} Cooperstock, F.I.: Found. Phys. {\bf 22}(1992)1011; in {\it Topics
in Quantum Gravity and Beyond: Papers in Honor of L. Witten} eds.
Mansouri, F. and Scanio, J.J. (World Scientific, Singapore, 1993)
201; in {\it Relativistic Astrophysics and Cosmology}, eds.
Buitrago et al (World Scientific, Singapore, 1997)61; Annals Phys.
{\bf 282}(2000)115.

\item{[5]} Misner, C.W., Thorne, K.S. and Wheeler, J.A.: {\it Gravitation}
(W.H. Freeman, San Francisco, 1973).

\item{[6]} Ehlers, J. and Kundt, W.: {\it Gravitation: An Introduction to Current
Research}, ed. L. Witten (Wiley, New York, 1962)49.

\item{[7]} Pirani, F.A.E.: {\it Gravitation: An Introduction to Current
Research}, ed. L. Witten (Wiley, New York, 1962)199.

\item{[8]} Weber, J. and Wheeler, J.A.: Rev. Mod. Phy. {\bf 29}(1957)509;\\
Weber, J.: {\it General Relativity and Gravitational Waves},
(Interscience, New York, 1961).

\item{[9]} Qadir, A. and Sharif, M.: Physics Letters {\bf A
167}(1992)331.

\item{[10]} Rosen, N: Helv. Phys. Acta. Suppl. {\bf 4}(1956)171.

\item{[11]} Scheidegger, A.E.: Rev. Mod. Phys. {\bf 25}(1953)451.

\item{[12]} Rosen, N.: Phys. Rev. {\bf 291}(1958)110.

\item{[13]} Rosen, N. and Virbhadra, K.S.: Gen. Rel. and Grav. {\bf
26}(1993)429.

\item{[14]} Virbhadra, K.S.: Pramana-J. Phys. {\bf 45}(1995)215.

\item{[15]} Brown, J.D. and York, J.w.: Phys. Rev. D {\bf 47}(1993)1407.

\item{[16]} Tolman, R.C.: {\it Relativity, Thermodynamics and Cosmology}
(Oxford University Press, Oxford).

\item{[17]} Landau, L.D. and Lifshitz, E.M.: {\it The Classical Theory
of Fields} (Pergamon Press, Oxford 1987).

\item{[18]} M$\ddot{o}$ller, C.: Ann. Phys. (NY) {\bf 4}(1958)347.

\item{[19]} Komar, A.: Phys. Rev. {\bf 113}(1959)934.

\item{[20]} M$\ddot{o}$ller, C.: Ann. Phys. (NY) {\bf 12}(1961)118;

\item{[21]} Bergqvist, G.: Class. Quantum Grav. {\bf 9}(1992)1753;
Bernstein, D.H. and Tod, K.P.: Phys. Rev. {\bf D 49}(1994)2808.

\item{[22]} Papapetrou, A.: Proc. R. Irish. Acad. {\bf A52}(1948)11.

\item{[23]} Weinberg, S.: {\it Gravitation and Cosmology:
Principles and Applications of General Theory of Relativity}
(John Wiley and Sons, Inc., New York, 1972).

\item{[24]} Tyron, E.P: Nature {\bf 246}(1973)396.

\item{[25]} Matos, T., Nunez, D. and Quevedo, H.: Phys. Rev. {\bf D 51}(1995)310.

\item{[26]} Matos, T., Nunez, D. and Rios, M.: Class.
Quantum Grav.{\bf 17}(2000)3917.

\item{[27]} E. Ayon-Beato and A. Garcia: Phys. Lett. {\bf B464}(1999)25.

\item{[28]} Tod, K. P.: Proc. R. Soc. London {\bf A388}(1983)457.

\item{[29]} Aguirregabiria, J.M., Chamorro, A. and Virbhadra, K.S.:
Gen Rel. Grav. {\bf 28}(1996)1393.

\item{[30]} Cohen, J.M. and de Felice, F.: J. Math. Phys. {\bf 25}(1984)992.

\item{[31]} Ernst, F.J.: J. Math. Phys. {\bf 15}(1975)1409; {\bf 17}(1976)54.

\item{[32]} Kramer, D., Stephani, H. , MacCallum, M. and Herlt, E.: {\it Exact
Solutions of Einstein's Field Equations} (Cambridge University
Press, Cambridge, 1980).

\item{[33]} Mashhoon, Bahram , McClune, James C. and Quevedo, Hernando: Class.
Quantum Grav.{\bf 17}(2000)533.

\item{[34]} Verdaguer, E.: Phys. Rep. {\bf 229}(1993)1;\\
Griffths, J.B.: {\it Colliding Plane Waves in General Relativity}
(Oxford University Press, Oxford, 1991).

\item{[35]} Alekseev, G.A. and Griffths, J.B.: Class. Quantum Grav.
{\bf 13}(1996)2191.

\item{[36]} Alekseev, G.A. and Griffths, G.B.: Class.
Quantum Grav.{\bf 12}(1995)L13.

\item{[37]} Qadir, Asghar, Sharif, M. and Shoaib, M.: Nuovo Cimento
{\bf B 13}(2000)419.

\end{description}

\end{document}